\documentclass[11pt]{article}
\usepackage{graphics,amssymb}
\usepackage{epsfig}
\begin{document}

{
\setlength{\textwidth}{16.5cm}
\setlength{\textheight}{22.2cm}
\setlength{\hoffset}{-1.43cm}
\setlength{\voffset}{-.9in}

\thispagestyle{empty}
\renewcommand{\thefootnote}{\fnsymbol{footnote}}

\begin{flushright}
{\normalsize
SLAC-AP-129\\
July 2000}
\end{flushright}

\vspace{.8cm}

\begin{center}
{\bf\Large On Resonant Multi-Bunch Wakefield Effects in Linear Accelerators
with Dipole Mode Detuning
\footnote{\small Work supported by
Department of Energy contract  DE--AC03--76SF00515.}}

\vspace{1cm}

{\large
Karl L.F. Bane and Zenghai Li\\
Stanford Linear Accelerator Center, Stanford University,
Stanford, CA  94309}

\end{center}
}
\vfill

\def\la{\langle} 
\def\ra{\rangle} 
\def\lm{\lambda}

\title{On Resonant Multi-Bunch Wakefield Effects in Linear Accelerators
with Dipole Mode Detuning}
\author{Karl L.F. Bane and Zenghai Li}
\date{}
\maketitle

In this report we explore resonant multi-bunch (transverse) wakefield
effects in a linear accelerator in which the dipole modes of the
accelerator structures have been detuned. 
For examples we will use the
parameters of a slightly simplified version of an
optimized S-band structure described in
Ref.~\cite{BL}. Note that we are also aware of a different analysis of
resonant multi-bunch wakefield effect\cite{schulte}.

It is easy to understand
how resonances can arise in a linac with
bunch trains. 
Consider first the case of the interaction
of the beam with one single structure mode.
The leading bunch enters the structure 
offset from the axis and excites the mode.
If the bunch train is sitting on an integer resonance,
{\it i.e.} if $f\Delta t=n$, with $f$ the mode frequency,
$\Delta t$ the bunch spacing, and $n$ an integer,
then when the 2nd bunch arrives it will excite the mode at the same phase
and also obtain a kick due to the wakefield of the first bunch.
The $m{\rm th}$ bunch will also excite the mode in the same phase
and obtain $(m-1)$ times the kick from the wakefield that the second bunch experienced
(for simplicity we assume the mode $Q$ is infinity).
On the half-integer resonance, {\it i.e.} when $f\Delta t=n+.5$, 
the $m{\rm th}$ bunch will also receive kicks 
from the wakefield left by the
earlier bunches, but in this case the kicks will
 alternate in direction, and no resonance builds up.
For a transverse wakefield effect, such as we are interested in
here, however,
this simple description of the resonant
interaction needs to be modified slightly.
For this case the wake varies as
$\sin(2\pi ft)$, and neither the integer nor
 the half-integer resonance condition 
will excite any wakefield for the following bunches. 
In this case resonant growth
is achieved at a slight deviation from 
the condition $f\Delta t=n$, as is shown below.

In the following, for simplicity,
 we will use the ``uncoupled'' model to 
investigate resonant effects in the sum wake for a structure with modes with a
uniform frequency distribution. 
According to this model
(see, for example, Ref.~\cite{Gluck})
\begin{equation}
W(t)\approx \sum^{N_c}_n 2 k_{sn}\sin({2\pi f_{sn}t/ c})
\quad\quad\quad[t\ {\rm small}],
\label{eqwakes}
\end{equation}
where $N_c$ is the number of cells in the structure,
and $f_{sn}$ and $k_{sn}$ are, respectively, the frequency
and kick factor at the synchronous point, for a periodic
structure with dimensions of cell $n$.
Therefore, one can predict the short time behavior of the wake without
solving for the eigenmodes of the system.
The point of using the uncoupled model is that it allows us to
study the effect of an idealized, uniform frequency distribution.
As is well known, an ideal (input) frequency distribution
becomes distorted by the cell-to-cell coupling of an accelerator structure.
(For simplicity we will drop the $s$ in the subscripts for frequency below.)
For examples we will use the
parameters of a slightly simplified version (all kick factors are equal,
the frequency distribution is uniform instead of trapezoidal) of the
optimized $3\pi/4$ S-band structure described in
Ref.~\cite{BL}: there are $N_c=102$ cells (also modes),
the central frequency ${\bar f}=3.92$~GHz, and the full-width of 
the distribution $\Delta_{\delta f}=5.8\%$;
for bunch structure
we consider the nominal configuration of $M=95$ bunches in a train and a 
bunch spacing $\Delta t=2.4$~ns.
The results for the real structure, with coupled modes, will be
slightly different yet qualitatively the same.

\begin{figure}[htb]
\centering
\epsfig{file=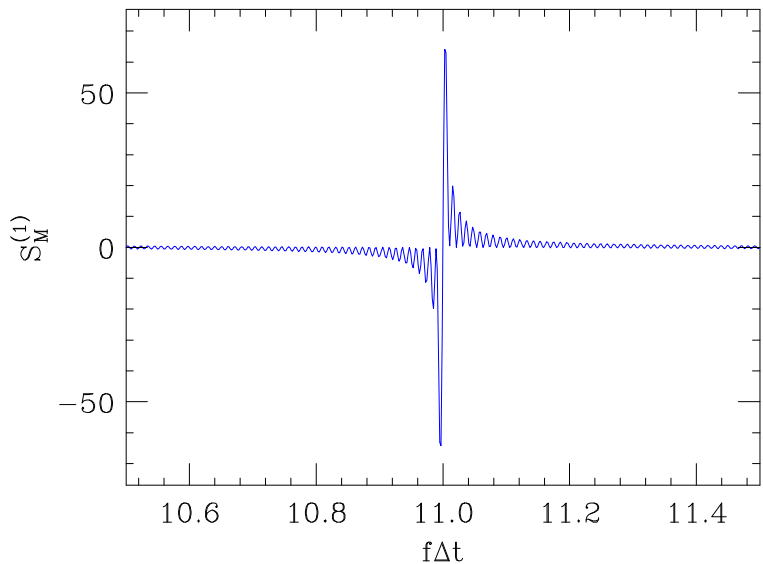, width=10cm}
\caption{
The sum wake at the last bunch in a train {\it vs} bunch spacing, due to a single mode
(Eq.~\ref{eqres1}); $M=95$ bunches.
}
\label{fires1}
\end{figure}

Consider first 
the case of a structure with only one dipole mode, with frequency $f$,
and a kick factor that we will normalize (for simplicity) to $1/2$.
Suppose there are $M$ bunches in the bunch train.
The sum wake at the $m{\rm th}$ bunch is given by
\begin{eqnarray}
S_m^{(1)}(f\Delta t)&=&\sum_{i=1}^m\sin\left(2\pi[i-1] f\Delta t\right)\nonumber\\
   &=& {\sin\left(\pi[m-1]f\Delta t\right)\sin\left(\pi mf\Delta t\right)
\over\sin\left(\pi f\Delta t\right)}\quad. 
\label{eqres1}
\end{eqnarray}
As with the nominal (2.8~ns) bunch spacing in the S-band prelinacs, let us,
for an example, consider $M=95$ bunches
and the region near the 11th harmonic. In Fig.~\ref{fires1}
 we plot $f\Delta t$ {\it vs}
the sum wake for the $M$th (the last) bunch, $S_M^{(1)}$, near the
11th integer resonance. 
It can be shown that, if $M$ is not small, the largest resonance peaks 
(the extrema of the curve) are at
\begin{equation}
f\Delta t\approx n\pm {3\over8M}\quad\quad[M\ {\rm not\ small}]\quad,
\end{equation}
with values $\pm.72M$.
Note that at the exact integer and half-integer resonant spacings the sum wake is zero.

\begin{figure}[htb]
\centering
\epsfig{file=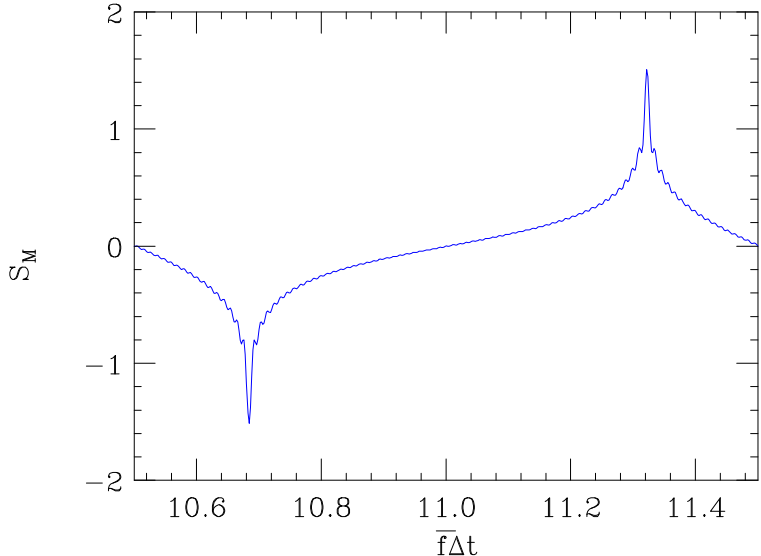, width=10cm}
\caption{
The sum wake at the last bunch in a train {\it vs} bunch spacing,
due to a uniform distribution of mode frequencies
(Eq.~\ref{equnib}).
 The total frequency spread $\Delta_{\delta f}=5.8\%$, and
$N_c=102$.
}
\label{fires2}
\end{figure}

Now let us consider a uniform distribution of mode frequencies.
For simplicity we will let all the kick factors be equal, and be normalized to
$1/2$. 
The sum wake, according to the uncoupled model, becomes
\begin{equation}
S_m({\bar f}\Delta t)={1\over N_c}\sum_{n=1}^{N_c}S_m^{(1)}\left[{\bar f}\Delta t\left(
1+{(n-N_c/2)\over N_c}\Delta_{\delta f}\right)\right]\quad,\label{equnib}
\end{equation}
with $N_c$ the number of cells (also the number of modes),
 $\bar f$ the central frequency, and $\Delta_{\delta f}$
the total (relative) width of the frequency distribution.
As an example, let us consider
the optimized $3\pi/4$ S-band structure, with $N_c=102$ and 
$\Delta_{\delta f}=5.8\%$. The sum wake at the last (the $M$th) 
bunch 
position, $S_M$, is plotted as function of ${\bar f}\Delta t$ in Fig.~\ref{fires2}.
Note that the uniform frequency distribution 
appears to suppress the integer resonance.
The extrema of the curve (the ``horns'') that are seen
at ${\bar f}\Delta t=11\pm.32$ 
are resonances due to the edges of the frequency distribution,
with the condition ${\bar f}\Delta t\approx11/(1\pm\Delta_{\delta f}/2)$.
Note, however, that the sizes of even these
spikes are small compared to those of the single mode case.

\begin{figure}[htb]
\centering
\epsfig{file=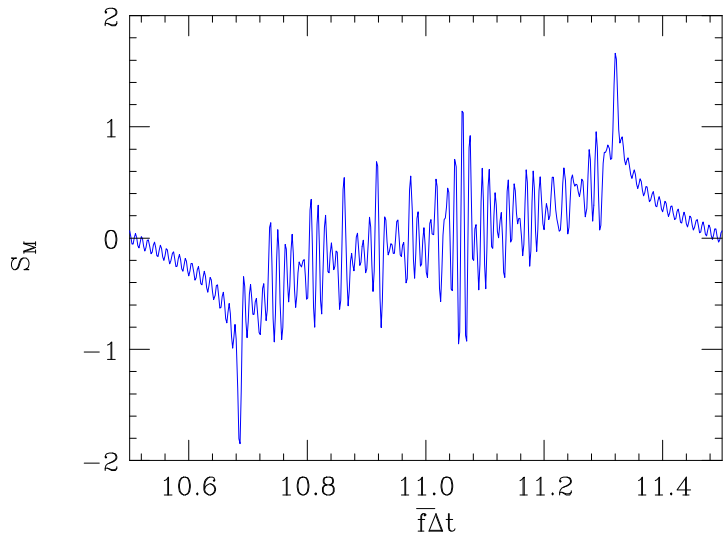, width=9.5cm}
\caption{
The sum wake at the last bunch in a train {\it vs} bunch spacing,
due to a uniform distribution of frequencies,
including frequency errors.
 The total frequency spread $\Delta_{\delta f}=5.8\%$, the number of
modes $N_c=102$, and rms relative frequency error is $10^{-4}$.
}
\label{fires3}
\end{figure}

\begin{figure}[htb]
\centering
\epsfig{file=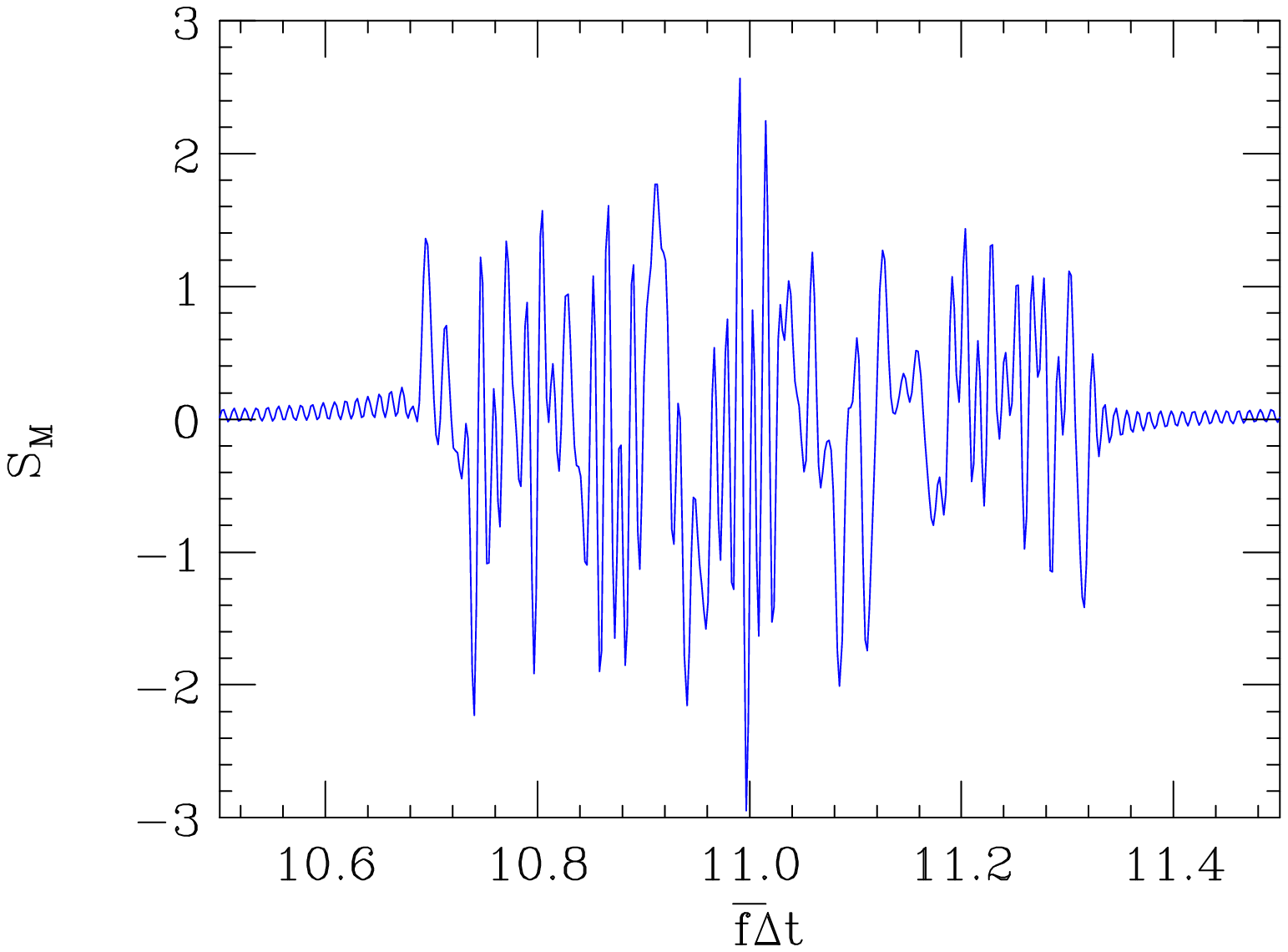, width=9.5cm}
\caption{
The sum wake at the last bunch in a train {\it vs} bunch spacing,
due to a uniform distribution of frequencies,
including random misalignment errors with rms 1.
The total frequency spread $\Delta_{\delta f}=5.8\%$ and then number of modes
$N_c=102$.
}
\label{fires4}
\end{figure}

Suppose we add frequency errors to our model. We can do this by, 
in each term in the sum of Eq.~\ref{equnib},
multiplying the frequency by the factor $(1+\delta f_{err}r_n)$,
 with $\delta f_{err}$ the rms (relative) frequency error and $r_n$ a random number
 with rms 1.
Doing this, considering a uniform distribution
in frequency errors with rms $\delta f_{err}=10^{-4}$,
 Fig.~\ref{fires2} becomes
Fig.~\ref{fires3}. 
Note that this 
perturbation is small compared to the frequency spacing
 $5.7\times10^{-4}$, so it does not really change the frequency distribution
significantly.
Nevertheless, because of 
resonance-like behavior 
we can see a large effect on $S_M$ throughout the range between the
horns of Fig.~\ref{fires2} ($10.68\leq {\bar f}\Delta t\leq 11.32$). 
To model cell-to-cell misalignments, we multiply
each term in the sum of Eq.~\ref{equnib} by the random factor $r_n$.
The results,
for a uniform distribution of 
errors with rms 1, are shown in Fig.~\ref{fires4}. Again resonance-like
behavior
is seen throughout the range between the horns of Fig.~\ref{fires2}. 

We can understand these results in the following manner:
Only when there are no errors 
does using a uniform frequency distribution
suppress the resonance in the region near the integer resonance.
But otherwise, using a uniform frequency distribution 
basically only reduces the size of the
resonances, at the expense of extending
the range in bunch spacings where they can be
excited. Instead of being localized in the region near the integer resonance
(${\bar f}\Delta t\approx n$), 
resonance-like behavior can now be excited anywhere between the limits
\begin{equation}
({\bar f}\Delta t)_\pm ={n\over1\mp\Delta_{\delta f}/2}\quad.
\end{equation}
Note that this implies that if $\Delta_{\delta f}>1/({\bar f}\Delta t)$,
then the resonance-like behavior 
cannot be avoided no matter what bunch spacing 
(fractional part) is chosen. For example, for the X-band linac in the NLC,
where the total width of the dipole frequency distribution 
(of the dominant first band modes) is 10\%,
even for the alternate (1.4~ns) bunch spacing, where the integer part
of ${\bar f}\Delta t$ is 21, the resonance region cannot be avoided.

\section*{Acknowledgments}

The authors thanks V.~Dolgashev for
carefully reading this manuscript.


\begin{thebibliography}{9}

\bibitem{BL}
K. Bane and Z. Li, ``Dipole Mode Detuning in the Injector Linacs
of the NLC,'' SLAC/LCC Note in preparation.

\bibitem{schulte}
D. Schulte, presentation given in an NLC Linac meeting, summer 1999.

\bibitem{Gluck}
K. Bane and R. Gluckstern, {\it Part. Accel.}, {\bf 42}, 123 (1994).


\end{thebibliography}
\end{document}